\documentclass[12pt]{article}

\usepackage{amsmath}
\usepackage{amssymb}

\def\rdots{\mathinner{\mkern1mu\raise1pt\vbox{\kern1pt\hbox{.}}\mkern2mu
   \raise4pt\hbox{.}\mkern2mu\raise7pt\hbox{.}\mkern1mu}}
\newcommand{\Z}{{\rm Z\kern-.35em Z}}
\newcommand{\bP}{{\rm I\kern-.15em P}}
\newcommand{\Q}{\kern.3em\rule{.07em}{.65em}\kern-.3em{\rm Q}}
\newcommand{\R}{{\rm I\kern-.15em R}}
\newcommand{\h}{{\rm I\kern-.15em H}}
\newcommand{\C}{\kern.3em\rule{.07em}{.65em}\kern-.3em{\rm C}}
\newcommand{\T}{{\rm T\kern-.35em T}}

\newcommand{\be}{\begin{equation}}
\newcommand{\ee}{\end{equation}}

\newcommand{\ve}{\varepsilon}
\newcommand{\la}{\lambda}
\newcommand{\Lam}{\Lambda}

\newcommand{\cd}{\cdot}
\newcommand{\ra}{\rightarrow}

\newcommand{\de}{\delta}
\newcommand{\Del}{\Delta}
\newcommand{\La}{\Lambda}

\newcommand{\al}{\alpha}
\newcommand{\nn}{\nonumber}

\newtheorem{lem}{Lemma}

\newtheorem{definition}{Definition}

\begin{document}
%\font\twelverm=cmr12
%\font\cs=CMSSI12

\openup 1.5\jot
\centerline{Tilings With Very Elastic Tiles}

\vspace{1in}
\centerline{Paul Federbush}
\centerline{Department of Mathematics}
\centerline{University of Michigan}
\centerline{Ann Arbor, MI 48109-1109}
\centerline{(pfed@umich.edu)}

\vspace{1in}

\centerline{\underline{Abstract}}

We consider tiles of some fixed size, with an associated weighting on the shapes of tile, of total mass 1.  We study the pressure, $p$, of tilings with those tiles; the pressure, one over the volume times the logarithm of the partition function.  (The quantity we define as ``pressure" could, perhaps equally harmoniously with physics notation, be called ``entropy per volume", neither nomenclature is ``correct".)  We let $\hat p^0$ (easy to compute) be the pressure in the limit of absolute smoothness (the weighting function is constant).  Then as smoothness, suitably defined, increases, $p$ converges to $\hat p^0$, uniformly in the volume.  It is the uniformity requirement that makes the result non-trivial.  This seems like a very basic result in the theory of pressure of tilings.  Though at the same time, perhaps non-glamorous, being bereft of geometry and not very difficult.  The problem arose for us out of study of a problem in mathematical physics, associated to a model of ferromagnetism.

\vfill\eject

\noindent
I.  \underline{Introduction}

We will be studying the partition function for tilings with tiles of some fixed size, with a weighting function on the tile shapes.  If the weighting is concentrated on a single tile shape, one for this shape, zero for all others, then one is computing the partition function for tilings with tiles all of this one shape.  If the weighting function assumes the value one on a set of $r$ shapes and is zero for every other shape, then the partition function just counts the number of tilings using tiles of these $r$ shapes.  (We will impose a normalization condition later, that would require the weighting on the $r$ shapes to each be $1/r$, a trivial change to accommodate.)  Our results will deal with weighting functions smoothly  varying functions of the tile shapes.  In this direction geometrical dependences will be seen to be unimportant.

We may think in terms of length scales, a favorite perspective of physicists.  The problem has three length scales:  1,  the length of the lattice bounds. $\ell$, the length scale characterizing the smoothness of the weighting function.   $L$, the edge length of the lattice cube.  In this language, we obtain limits on the pressure as $1/\ell$ becomes small.  The limits do not require $L/\ell$ to be small!  (This is the uniformity required later, in $|\Lambda|.$)  This paper works with a particular concept of smoothness defined below.  One direction for further work is considering other weaker definitions for smoothness.

It should be straightforward to consider tilings with different size tiles and to generalize the result herein to this situation.  Here we would let the mass of the weightings be different for different size tiles.  Then as the weighting function became smoother the pressure would approach a value depending only on the masses, the convergence uniform in lattice size.  Statistical mechanics applications might involve similar theorems for partial tilings, tilings that cover a certain fixed fraction of the lattice.  Much more general results than the present should be possible.

\noindent
II.  \underline{Definitions and Results}

We work with a $d$-dimensional cubic lattice, $\Lambda$, taken to be periodic.  (But our results will be independent of boundary properties of the lattice.)  We let $N$ be the number of vertices of $\Lambda$.

The {\it tiles} we consider will be of single {\it size} $n$.  They will not be required to be connected.  Tiles of size $n$ are identified with subsets of $n$ vertices of $\La$, under the equivalence relation that two subsets are equivalent if one is a translation of the other.

An {\it activity} (or {\it weighting}) on the tiles is a symmetric function on $\La^n, f(i_1,...,i_n),$ (with the same periodicity as $\La$) that satisfies
\begin{eqnarray}
&1)&  f(i_1,...,i_n) > 0, \ \ {\rm positivity}. \\
&2)&  f(i_1 +c, \; i_2+c,...,i_n+c) = f(i_1,...,i_n) \ \ {\rm translation \ invariance}. \\
&3)&  \frac{1}{(n-1)!} \, \displaystyle{\sum_{i_2,..., i_n}}f(i_1,i_2,...,i_n) = 1 \ \ {\rm normalization}.
\end{eqnarray}
In the sum in 3) it is understood $i_1,...,i_n$ must be distinct.  The strong positivity condition of 1) implies every possible shape tile is included in our set of tiles.  The values at the diagonal points are not ``physical" but will be involved later in equation (4).

We define $sm(f)$, the first parameter measuring the smoothness of $f$, by letting $sm(f)$ be the smallest constant such that

\be	|f(\vec{i} + \vec{u}) - f(\vec{i})| \le sm(f) \cd f(\vec{i})	\ee
for all $\vec{i} = i_1,...,i_n$ and $\vec{u}$ unit lattice vectors.   We include in our definition of smoothness the requirement that there be an $R$ such that
\be |f(i_1,...,i_n)| \le e^{-\max_{k,j}(|i_k - i_j)/R}	\ee
$R(f)$ is the smallest such $R$.  Thus smoothness is measured by two parameters $sm(f)$ and $R(f)$; our definition of smoothness has then a ``localization" included in it.

We do not believe that the localization requirement, equation (5), is necessary for our results, but the proof would be much harder along lines we have considered.  Here certainly is a path for future investigation.

We always assume $n$ divides $N$ and we define the {\it partition function} associated to the activity $f$ on tiles of $\Lam$ as the sum over tilings of $\Lam$ (by tiles of size $n$), each tiling weighted by the product of the activities of the individual tiles in the tiling.  This we denote by
\be	Z = Z(f) = Z(f, \Lam) \equiv e^{Np}	\ee
\be   p = p(f) = p(f, \Lam)	\ee
introducing here the {\it pressure}, $p$.

If $sm(f) = 0$, $f$ is a constant function, and it is easy to compute the corresponding partition functions, $\hat Z^0$.  $\hat Z^0$ depends only on $n$ and $N$ is given by
\be
\hat Z^0 = \left( \frac{(n-1)!(N-n)!}{(N-1)!} \right)^{N/n} \ \cdot \ \left( \frac{N!}{(\frac N n)! \ (n!)^{N/n}} \right)	\ee
as computed in Section VI.  The infinite volume limit for the pressure is a trivial consequence of (8).  With
\be	\hat Z^0 = e^{N\hat p^0}	\ee
one gets
\[	\lim_{N \ra \infty} \hat p^0 = \left( \frac{1-n}n \right),	\]
and we set 
\be	Z^0 \equiv e^{(\frac{1-n}{n})}  \ee

The following theorem is the goal of this paper.

\bigskip

\centerline{- - - - - - - - - - - - - - - - - - - -}

\bigskip
\noindent
\underline{Main Theorem}  For each $\ve > 0$ and $c_1 > 1$ there is a $\de = \de(\ve, c_1)$ such that 
\be	|p(f,\Lam) - \hat p^0 (n,N)| < \ve	\ee
if
\be sm(f) < \de  \ee
and
\be R(f) < c_1/\de .	\ee

\bigskip

\centerline{- - - - - - - - - - - - - - - - - - - -}

\bigskip

The form of the conditions (12) and (13) might be motivated by considering scaling of the function $f$ 
\be  f^\la(\vec{x}) \equiv \la^df(\la {\vec x}) \ee
and noting how $sm$ and $R$ change (true, $f^\la$ will not exactly satisfy the normalization condition, equation (3)). 

We note again that we expect the condition (13) of the theorem to be unnecessary.  It is the requirement that $\de$ not depend on $|\Lam|$ that makes the theorem non-trivial.

\noindent
III.  \underline{Proofs I, Preliminaries}

\begin{lem}.  There is a universal upper bound $M$, on all the partition functions we deal with
\be Z^{1/N} \ \ \le \ \ M .  \ee
\end{lem}

\bigskip

\centerline{- - - - - - - - - - - - - - - - - - - -}

\bigskip

It is easy to see that 
\be	M = \left( Z^+ \right)^{1/N}	\ee
works, where $Z^+$ is given by Definition 1 and equation (34) of the next section, with $\bar{f} = f$ and $\bar n = 1$.  This best $M$ depends on $N$ and $n$, but a less ideal bound may be chosen a function only of $n$.  The best  $M$, from (34) as just specified is explicitly
\be \left[ \frac 1{(N/n)!} \ \left( \frac 1 n \right)^{N/n} \ N^{N/n} \right]^{1/N}.		\ee
The factors in this equation are explained by considering the sum over each $\cal S$ in (34).  We pick a preferred element in $\cal S$, and then sum over the complementary sets of $n-1$ elements.  The sum over the location of the preferred element yields a factor of $N$.  Then we divide by $n$ to correct for the over counting involved in selecting the preferred element.

\bigskip

\centerline{- - - - - - - - - - - - - - - - - - - -}

\bigskip

\begin{lem}.  There is a $c$ such that if for two weighting functions $f_1$ and $f_2$ satisfying (1), (2) and (3) one has
\be	|f_1- f_2| \le \ve \; f_1	\ee
then
\be \Big | \left( Z(f_1) \right)^{1/N} - \left( Z(f_2) \right)^{1/N} \Big| \le c\;\ve \ . \ee
\end{lem}

\bigskip

\centerline{- - - - - - - - - - - - - - - - - - - -}

\medskip
\noindent
Equation (18) is a pointwise bound.  This lemma is a consequence of the following estimate.

\medskip
\centerline{- - - - - - - - - - - - - - - - - - - -}
\medskip

\noindent
\underline{Root Estimate}.  Let $a_{ij} \ge 0$ and $|\de_{ij}| \le 1$.  Set
\be	\bar\de = \max (\{ \de_{ij}\})	\ee
and
\be	\hat\de = \min (\{ \de_{ij}\})	\ee
and define
\be	A = \left( \sum_i \prod^N_{j=1} a_{ij} \right)^{1/N} \ . \ee
Then
\be	(1 - |\hat\de|)A \le  \left( \sum_i \prod^N_{j=1} a_{ij}(1+\de_{ij}) \right)^{1/N} \le (1 + |\bar \de|)A \ . \ee 

\medskip
\centerline{- - - - - - - - - - - - - - - - - - - -}
\medskip

\noindent
It is a simple matter to write the root of the partition function $Z(f_1))^{1/N}$ in the form of (22) and  $Z(f_2))^{1/N}$ as the intermediate term in (23).  One also uses Lemma 1 to write the bound in the form (19).

We will find it convenient to divide cube $\Lambda$ of edge size $L$ into $\bar N$ smaller cubes of edge size $\bar{\ell}$.  We have the following relations under this subdivision, introducing the natural quantities $\bar L$ and $\bar n$.

\begin{eqnarray}
|\Lambda| &=& L^d = N \\
L &=& \bar L \; \bar{\ell} \\
\bar{\ell}^d &=& \bar n \\
\bar N &=& N/\bar n = \bar L^d
\end{eqnarray}
Of course we must assume $\bar{\ell}$ divides $L$.

We set $c_i$ to be the set of vertices in the $i^{th}$ side $\bar{\ell}$ little cube in this dissection.  We define now $\bar f$, some sort of averaging of $f$ over the little cubes.  $\bar f(i_1,...,i_n)$ will depend only on which  $c_k$ each $i_j$ is in.  Suppose
\be    i_r \in c_{g(r)} \ \ \ \ r = 1,...,n \ .  \ee
Then we set
\be	\bar f \Big(i_1,..., i_n\Big) = \frac 1 {\cal N} \sum_{\ell_1 \in c_{g(1)}} \cdots \sum_{\ell_n \in c_{g(n)}} \; f \Big(\ell_1,..., \ell_n\Big) \ee
where $\cal N$ is the number of vertices in
\[	c_{g(1)} \times c_{g(2)} \times ... \times c_{g(n)}   \]
that are off the diagonals, and the $\ell_i$ in the sum in (29) are distinct.  (Where we use (29) $\bar n$ will always be greater than $n$.) 

We now present a pointwise bound on the difference between $f$ and $\bar f$, arising directly from equation (4).

\medskip
\centerline{- - - - - - - - - - - - - - - - - - - -}
\medskip

\begin{lem}.   If
\be		\alpha \equiv \bar{\ell} dn \ sm(f) < 1	\ee
then
\be		|f - \bar f| \le \frac {\al}{1-\al} \ f		\ee
\end{lem}

\medskip
\centerline{- - - - - - - - - - - - - - - - - - - -}
\medskip
 
Considering a single polycube as in (29), we let $f_m$ and $f_M$ be the minimum and maximum values of $f$ on this polycube.  Being slightly schematic in our notation, we apply (4) to a shortest path in the polycube connecting the vertices where $f$ achieves the values $f_m$ and  $f_M$ 
\begin{eqnarray}
|f_m - f_M| &\le& \sum_k \Delta_k \; f \le nd\bar{\ell} \ \max (\Del_k f) \\
&\le& nd\bar{\ell}\; sm(f) f_M \le nd\bar{\ell}\; sm(f) \frac{f_M}{f_i} \ f_i \nn \\
&\le& nd\bar{\ell} \; sm(f) \ \frac{f_M}{f_M} \ f_i \nn \\
&\le& \frac\al{1-\al} \ f_i \nn
\end{eqnarray}
where $f_i$ is value of $f$ at any vertex in the polycube.

\bigskip
\bigskip

\noindent
IV.  \underline{Proofs 2, the Idea}.

A subset, $\cal S$, of the vertices of $\La$ with $\#({\cal S}) = n$ determines a tile placed on a particular place on the lattice, a tile with weighting $f(\cal S)$.  If $f$ is the symmetric function on $\La^n$ determining a weighting, $f$ is  canonically a function on such $\cal S$.

\medskip
\centerline{- - - - - - - - - - - - - - - - - - - -}
\medskip

\begin{lem}.
\be	\sum_{\cal S} \ f({\cal S}) = N/n.  \ee
\end{lem}

\noindent
\underline{Proof}.  Picking one of the elements of $\cal S$ it may be placed in $N$ different positions.  The sum of $f$ over the remaining elements of $\cal S$ is one, by the normalization condition (3).  But there were $n$ choices for the initial element picked, so we are overcounting by a factor of $n$.

\medskip
\centerline{- - - - - - - - - - - - - - - - - - - -}
\medskip

We continue working with a particular dissection into smaller cubes, with the notation of (24) - (27) and the corresponding $f$ and $\bar f$ weighting functions.  We associate to $\bar f$, in addition to the partition function, $Z(\bar f)$, three other ``approximate" partition functions, $Z^+(\bar f), \ Z'(\bar f)$, and $Z^-(\bar f)$.  All three of these partition functions arise from the following single proto-equation.

\medskip
\centerline{- - - - - - - - - - - - - - - - - - - -}
\medskip

\be Z^\bullet (\bar f) \equiv \frac 1{(N/n)!} \cdot \sum^\bullet_{ {\cal S}_1,{\cal S}_2,...,{\cal S}_{N/n} } \Pi_i \bar f({\cal S}_i) \cdot \left( \frac{\bar n !}{\bar n^{\bar n}} \right)^{{\bar N}} .   \ee
Here the $\bullet$  over the sum indicates specific restrictions on the sum.

\medskip
\centerline{- - - - - - - - - - - - - - - - - - - -}
\medskip

\begin{definition}.  $Z^+(\bar f)$ is defined by equation (34) with no restrictions on the sum.
\end{definition}

\medskip
\centerline{- - - - - - - - - - - - - - - - - - - -}
\medskip

As above, we let $c_i$ be the set of vertices in the $i^{th}$ side $\bar{\ell}$ little cube in the dissection of (24) - (27).  For a given term of the sum in (34), we set
\be	\tilde{n}_i = \sum_j \ \# \{{\cal S}_j \cap c_i \} \ . \ee
$\tilde{n}_i$ is thus the number of tile vertices that fall in $c_i$, for a true tiling $\tilde{n}_i$ must equal $\bar n$.

\medskip
\centerline{- - - - - - - - - - - - - - - - - - - -}
\medskip

\begin{definition}.  $Z'(\bar f)$ is defined by equation (34) above, with the restriction that the terms kept in sum are  exactly those such that $\tilde{n}_i =\bar n$, all $i$.
\end{definition}

\medskip
\centerline{- - - - - - - - - - - - - - - - - - - -}
\medskip

\begin{definition}.   $Z^-(\bar f)$ will be defined explicitly in the next section.  It will be given as a sub sum of terms kept in $Z'(\bar f)$.
\end{definition}

\medskip
\centerline{- - - - - - - - - - - - - - - - - - - -}
\medskip

In proving the Main Theorem (see (11) - (13)) we will for each $\ve, c_1$, and $sm(f)$, as in the statement of this theorem, specify an $\bar{\ell}$ and thus have an $\bar f$ associated to $f$.  The following steps, stated in capsulized form, will yield the proof of the Main Theorem. 

\medskip
\centerline{- - - - - - - - - - - - - - - - - - - -}
\medskip

\noindent
\underline{Step 1}

\be \left(Z^+(\bar f)\right)^{\frac 1 N} \ge \left(Z'(\bar f)\right)^{\frac 1 N} \ge \left(Z^-(\bar f)\right)^{\frac 1 N} \ee

\bigskip

\noindent
\underline{Step 2}

\be  \left(Z(\bar f)\right)^{\frac 1 N} \longrightarrow \left(Z (f)\right)^{\frac 1 N}  \ee

\noindent
\underline{Step 3}

\be  \left(Z^+(\bar f)\right)^{\frac 1 N} \longrightarrow \left( Z^0 \right)^{\frac 1 N}  \ee

\noindent
\underline{Step 4}

\be  \left(Z^-(\bar f)\right)^{\frac 1 N} \longrightarrow \left(Z^+(\bar f) \right)^{\frac 1 N}  \ee

\medskip
\centerline{- - - - - - - - - - - - - - - - - - - -}
\medskip

Relation (36), i.e. Step 1, follows immediately from the Definitions 1,2,3 above.   Lemma 2 and Lemma 3 are all we will need to interpret relation (37) of Step 2.  We will leave till the next section specification of the exact sense of the limits in (37) - (39).  $Z^0$ in (38) is from equation 10.  Step 3 is the Theorem B of the next section, and Step 4 Theorem C.  Step 1 and Step 4 will imply
\be	\left( Z'(\bar f)\right)^{\frac 1 N} \longrightarrow \left( Z^+(\bar f)\right)^{\frac 1 N} \ee
and then from Step 3, (38)
\be	\left( Z'(\bar f)\right)^{\frac 1 N} \longrightarrow \left( Z^0\right)^{\frac 1 N} \ .\ee
\noindent
\underline{Step 5}
\be	\left( Z'(\bar f)\right)^{\frac 1 N} \longrightarrow \left( Z(\bar f)\right)^{\frac 1 N} \ .\ee 
This will be Theorem A of the next section.  Putting all these steps together we easily see all the $Z's$ $Z^+, Z', Z^-, Z^0, Z$, become equal in the limit.

\bigskip
\bigskip
\noindent
V.  \underline{Proofs 3, the Nitty-Gritty}.

As in the discussion after Definition 3, we note in the set up of the Main Theorem, for each $\ve, c_1$, and $sm(f)$, we will later associate an $\bar n$ and thus an $\bar f$.  Now we just note that as $\ve \ra 0$ one will have $\bar n \ra \infty$.  This will ensure that the following form for Theorem A suffices to explicit Step 5, equation (42) 

\medskip
\centerline{- - - - - - - - - - - - - - - - - - - -}
\medskip

\noindent
\underline{Theorem A}  Given $\ve > 0$, there is an $n(\ve)$ such that
\be	|\left( Z'(\bar f)\right)^{\frac 1 N} - \left( Z(\bar f)\right)^{\frac 1 N}| < \ve \ee
if $\bar n > n(\ve)$ 

\medskip
\centerline{- - - - - - - - - - - - - - - - - - - -}
\medskip

In all our proofs we will assume that the value of $\bar n$ divides the value of $N$.  Alternatively we could eliminate this requirement and have some unequal-sided boxes along the boundary.  With a little more work one can prove the same theorems in the latter case.

\medskip
\centerline{- - - - - - - - - - - - - - - - - - - -}
\medskip

\noindent
\underline{Proof}.  The sum over each $\cal S$ is discussed in connection with Lemma 1 of the last section.  The factor of $(\frac {\bar n \; !}{\bar n^{\bar n}})^{\bar N}$ in (34) would exactly take care of the fact that we have allowed vertices to overlap within each little cube (from the special property of $\bar f$) ..... {\it except} that in the sum for the vertices  of each individual $\cal S$ vertices are not allowed to overlap.  Thus we are accounting for not overlapping twice.  Say ${\cal S}_k$ has $r$ vertices lying in box $c_j$.  The sum over these $r$ vertices would have a factor $\bar n \; !\big/ (\bar n - r)!$.  The corresponding contribution as given by (34) would be correct if this were instead $\bar n^r$.  We then may use the root estimate, eq. (23), to yield the proof. 

\medskip
\centerline{- - - - - - - - - - - - - - - - - - - -}
\medskip

\noindent
\underline{Theorem B}.  Given $\ve > 0$, there are $N(\ve)$ and $n_1(\ve)$ such that

\be	|\left( Z^+\right)^{\frac 1 N} - \left( Z^0\right)^{\frac 1 N}| \le \ve  \ee
if $\bar n > n_1(\ve)$ and $N > N(\ve)$.

\medskip
\centerline{- - - - - - - - - - - - - - - - - - - -}
\medskip

The explicit expression for $Z^+(\bar f)$ is
\be	\frac 1{(\frac N n )!} \cdot N^{N/n} \cdot \left(\frac {\bar n \; !}{\bar n^{\bar n}}\right)^{\bar N} \cdot \left( \frac 1 n \right)^{N/n} . \ee
The last factor arises from the sum over $\cal S$'s as described in the proof of Lemma 1 (see equation (17)), as does the factor of $N^{N/n}$.  Equation (44) follows easily using Stirling's formula
\be ln \ s! = slns - s + {\cal O}(s) . \ee

\medskip
\centerline{- - - - - - - - - - - - - - - - - - - -}
\medskip

We now begin the rather long trek to a proof of Theorem C.  This is the difficult part of the paper.

\medskip
\centerline{- - - - - - - - - - - - - - - - - - - -}
\medskip

\noindent
\underline{Construction of $Z^-(\bar f)$}.

We will use some new notation.  In addition to a {\it set}, $\cal S$, a set of $n$ distinct vertices in $\La$, say
\be	i_1, i_2, \dots, i_n \longleftrightarrow \cal S, 	\ee
we will deal with a {\it pointed set}, ${\cal S}^\bullet$, which is a set of $n$ distinct vertices with one distinguished, which we pick to be the first in the ordering.  We may say from equation (3) that the sum
\be \sum \ f({\cal S}^{\bullet}) = 1	   \ee
where the sum is over all (pointed) sets {\it pointed} at the same point.

We now define {\it supersets} (which will all be pointed).  A {\it superset} $\bar {\cal S}$, is a sequence of small cubes (as in the discussion around (29))
\be 	\bar{\cal S} \longleftrightarrow c_1, c_2, ...., c_n.	\ee

Repetitions are allowed, and the superset is pointed at $c_1$.  We identify supersets $\bar{\cal S}_1  \longleftrightarrow (c_1,...,c_n)$ and $\bar {\cal S}_2 \longleftrightarrow (d_1,...,d_n)$ if $c_1 = d_1$ and $c_2,...,c_n$ is a permutation of $d_2,...,d_n$.  A pointed set ${\cal S}^{\bullet}$ is {\it covered} by $\bar{\cal S}$ of (49) if with
\[	{\cal S}^{\bullet} \longleftrightarrow (i_1,...,i_n)   \]
one has 
\[	i_1 \in c_1	\]
and there is a permutation $p$ of $2,...,n$ such that
\[	i_s \in c_{p(s)} \ \ \ \ \ s = 2,...,n \ .   \]

Two supersets $\bar{\cal S}_1$ and $\bar{\cal S}_2$ 
\begin{eqnarray*}
\bar{\cal S}_1 &\longleftrightarrow& c_1,...,c_n \\
\bar{\cal S}_2 &\longleftrightarrow& d_1,...,d_n 
\end{eqnarray*}
are of the same {\it supertype}, $\cal T$ if there is a permutation $p$ of $2,...,n$ such that
\[	c_1, c_{p(2)},...,c_{p(n)}  \]
is a translation of
\[	d_1,...,d_n  \]
i.e. there is a vector $\vec c$ such that 
\[	c_1  = d_1 + \vec c   \]
and
\[	c_{p(j)} = d_j + \vec c, \ \ \ \ j = 2, \dots, n.	\]
We considered giving supertypes the name {\it pointed supertiles}.

Now to defining $Z^-(\bar f)$ by detailing the sum in equation (34).  We will sum over pointed sets instead of sets, and later correct for the overcounting.  We first sum over the vertices at which each of the $N/n$ sets is pointed.  We restrict the sum to the situation where there is an equal number of such vertices, $\bar n/n$, in each little cube.  Let $z_{i1},\dots,z_{i \; \bar n/n}$ be the $\bar n/n$ such vertices in cube $c_i$.  To each of these is associated a superset pointed at $c_i$ say $\bar{\cal S}_{ij}$ to $z_{ij}$.  A remaining sum is over pointed sets for each $z_{ij}$ , the sets pointed at $z_{ij}$ covered by $\bar{\cal S}_{ij}$.

Let the index set I enumerate all supertypes.  We assume assigned to supertype ${\cal T}_k$, $k\in I$, a number $\al_k$ satisfying
\begin{eqnarray}
\al_k &\ge& 0 \\
\sum \; \al_k &=& 1
\end{eqnarray}
and 
\be \al_k (\bar n/n) \ \  {\rm is \ an \ integer} .\ee
 
We then require that in the assignment of supersets to each $c_i$ that $\al_k(\bar n/n)$ are of type ${\bar T}_k$ for each $k$ in $I$.  Our construction of $Z^-(\bar f)$ assumes a particular selection of $\al_k$, one we later specify.  One should visualize the construction long enough to realize in our sum all terms have $\tilde n_i = \bar n$ all $i$, see(35). 

\noindent
\underline{Computation of $Z^-(\bar f)$}.

In fact we compute a lower bound for $Z^-(\bar f)$, a product of factors $F_1,...,F_5$ we derive below.  (Veterans of the cluster expansion campaigns of CQFT can at least count.)  We consider $F_i$ in sequence.

1)  \underline{Factors explicit in equation (34)}.

We include these in $F_1$.
\be	F_1 = \frac 1{(\frac N n )!} \left( \frac {\bar n \; !}{\bar n^{\bar n}}\right)^{\bar N} . \ee

2)  \underline{Correcting for the overcounting}.

As mentioned before, by using pointed sets we are overcounting; each set may be pointed in $n$ different ways.  So we set $F_2$ to correct for the overcounting
\be	F_2 = \left( \frac 1 n \right)^{N/n} .  \ee
\underline{But}, in our construction of $Z^-$ we restrict ourselves to choices of the pointed sets such that exactly $\bar n/n$ are pointed in each little cube.  So not all the overcounting is actually done.  Thus in multiplying by $F_2$ we will get an underestimate for $Z^-$.

3)  \underline{Picking the pointing locations, the} $z_{ij}$.

\be F_3 = \frac{(N/n)!} {\left( (\bar n/n)!\right)^{N/\bar n}} \cdot \ \bar n^{N/n} \ee

The first factor counts the number of ways of putting the $(N/n)$ pointing locations into the $N/\bar n$ boxes, $\bar n/n$ in each box.  The second factor locates each pointing within a box.

4)  \underline{Assigning supertypes to the $z_{ij}$}.

In box $c_i$ there are $\al_k(\bar n/n)$ supertypes of type $k$.  The number of ways of making such an assignment to the $z_{ik}$, with $i$ fixed, is
\be \frac{(\bar n/n)!}{\Pi_k (\al_k (\bar n/n))!} \ . \ee

So for all boxes we get
\be   F_4 = \left( \frac{(\bar n/n)!}{\Pi_k \; (\al_k ( \bar n/n))!} \right)^{N/\bar n} \; . \ee

5)  \underline{Sum over pointed sets}

Let $\bar{\cal S}$ be a superset pointed at $c_i$ of type $k$.  We define
\be	a_k \equiv \sum \; \bar f({\cal S}^\bullet )	\ee
where the sum is over ${\cal S}^\bullet $  pointed at some fixed point in $c_i$, with ${\cal S}^\bullet$ covered by $\bar{\cal S}$.  It follows from equation (3) that
\be	\sum_k \; a _k = 1.   \ee
With those $a_k$ defined we find
\be		F_5 = \left( \Pi_k \; a_k ^{\al_k (\bar n/n)} \right)^{N/\bar n} \ee

\bigskip
\bigskip

\underline{The Why and Wherefore}.

Within the next few lines we will see the motivation for the construction of $Z^-$, find out the approximate values to be chosen for the parameters $\al_k$, and anticipate the framework for the proof of Theorem C.  Collecting factors $F_1,...,F_5$ we have now
\begin{eqnarray}
\left(Z^-(\bar f)\right)^{1/N} &\ge& (F_1 F_2 .... F_5)^{1/N} \\
&\ge& \left( \frac {\bar n !}{\bar n^{\bar n}}\right)^{1/\bar n} \left( \frac{\bar n}n \right)^{1/n} 
\left( \Pi_k \ \frac{ a_k^{\al_k ( \bar n/n)}} {(\al_k(\bar n/n))!} \right)^{1/\bar n} \ .
\end{eqnarray}
If each factorial in this expression were replaced by the approximation
\be	r! \ \approx \ (\frac r e )^r	\ee
then (62) would become
\be	\gtrsim \ e^{\left( \frac{1-n}n \right)} \ \Pi_k \left( \frac{a_k}{\al_k}\right)^{(1/n ) \al_k} .	\ee
Picking 
\be	\al_k = a_k	\ee
the right side of (64) achieves its maximum
\be Z^-(\bar f) \gtrsim e^{\left(\frac{1-n}n \right)} \ee
Compare equations (9) and (10).  To prove Theorem C we must adjust the $\al_k, N, \bar n$ values so that (66) is almost correct (depending on $\ve$).

Given $\ve > 0$ we will find conditions of the form (12) and (13) that guarantee
\be |\left(Z^+(\bar f)\right)^{1/N} - \left(Z^0 \right)^{1/N}| < \ve /2	\ee
and
\be  \left(Z^-(\bar f)\right)^{1/N} > e^{\frac{1-n}n} - \ve/2	\ee
from which
\be |\left(Z^+(\bar f)\right)^{1/N} - \left(Z^-(\bar f) \right)^{1/N}| < \ve 	\ee
which will yield Theorem C and the Main Theorem.  This will proceed through a number of steps, and deal with conditions that are not always in the form of (12) and (13).

\medskip
\centerline{- - - - - - - - - - - - - - - - - - - -}
\medskip

\noindent
\underline{Step 1}.  There are $N_1$ and $\bar n_1$ such that
\be |\left(Z^+(\bar f)\right)^{1/N} - \left(Z^0 \right)^{1/N}| < \ve /2	\ee
if $N > N_1$ and $\bar n > \bar n_1$.  This is Theorem B, equation (44).

\medskip
\centerline{- - - - - - - - - - - - - - - - - - - -}
\medskip

We rewrite equation (62) in the following form exhibiting Stirling's formula terms to be controlled. 
\be 
\left(Z^-(\bar f)\right)^{1/N} \ge e^{ \frac{1-n}{n} }
 \cdot \left( \frac {\bar n !}{(\bar n /e)^{\bar n}} \right)^{1/\bar n}  
\cdot  \Pi_k \ \left( \frac{( \frac{\al_k\bar n}{ne})^{\al_k (\bar n/n)}} {(\al_k (\bar n/n))!} \right)^{1/\bar n} \ \cdot
\ \Pi_k \left( \frac{a_k}{\al_k} \right)^{\al_k/n} .
\ee
We will use upper and lower bounds for Stirling's formula
\be		(2\pi r)^{1/2} e^{ \frac{1}{12 r}}  > \frac{r !}{(r/e)^r} > (2\pi r)^{1/2}	\ee
from Section 2.7 of Feller's book [1].  For the second factor in (71) we need only note
\be	\left( \frac {\bar n !}{(\bar n /e)^{\bar n}} \right)^{1/\bar n} > 1 .  \ee

\medskip
\centerline{- - - - - - - - - - - - - - - - - - - -}
\medskip

\noindent
\underline{Step 2}.  We arrange the $a_k$ in some order and let $\bar M$ be such that
\be	\sum^{\bar M}_{k=1} \ a_k > 1 - \frac {\ve}{20} . \ee
We can order this sum so that
\be		a_K > \frac{\ve}{20\bar M} \ \ \ \ {\rm  if} \ \ \ \ \ \ k < M   \ee
and
\be    \sum^M_{k=1} \ a_k > 1 - \frac \ve{10} \ .    \ee
In Section VI we will find relations between $sm(f)$, $\bar n$, $R(f)$, and $\bar M$.  We will set
\be		\al_k = 0 \ \ \ \ {\rm if} \ \ \ \ k > M.	\ee

\medskip
\centerline{- - - - - - - - - - - - - - - - - - - -}
\medskip

\noindent
\underline{Step 3}.  We would like to pick
\be		\al_k = \frac{a_k}{\sum^M_{k=1}a_k} \ \equiv \ \bar\al_k		\ee
but the $\al_k$ must satisfy
\be		\al_k \ \left( \frac{\bar n}{n} \right) \ \ {\rm is \ an \ integer}	\ee
by (52).  We will settle for
\be |\al_k - \bar \al_k| \le \frac \ve{10} \ a_k .	\ee

\medskip
\centerline{- - - - - - - - - - - - - - - - - - - -}
\medskip

\begin{lem} If
\be			\bar n \ > \ \frac{200\; \bar M}{\ve^2} \ n		\ee
then we may find $\al_k$ satisfying (80).  Always we impose
\end{lem}
\be	\sum \ \al_k = 1.	\ee

\noindent
\underline{Proof}.  By (75) it is enough to prove we may find $\al$'s satisfying
\be		|\al_k - \bar \al_k| \le \frac{\ve^2}{200} \ \bar M \ ,	\ee 
assume $\frac n{\bar n}$ satisfies (81).  On the real line lay down the lattice points that are integral multiples of $\frac n{\bar n}$ and also the points $\bar \al_k$.  If all the $\bar \al_k$ lie on the lattice points we may pick $\al_k =\bar \al_k$, satisfying (79) and we are done.  Otherwise we start out with $\al_k = \bar \al_k$, and then move the $\al_k$ points while satisfying (82).  If we can thus move the $\al_k$ till they all lie on the lattice points, {\it while never moving an $\al_k$ outside the interval between lattice points it lies within}, we will have proved the lemma.  The final position of $\al_k$, lying on a lattice point, satisfies (79).  Not leaving the interval it starts within, ensures (83).  We move the $\al$'s two at a time, satisfying (82), always moving an $\al$ closest to the lattice point to the left of it and an $\al$ closest to the lattice point to the right of it (neither point on the lattice), towards the lattice points.  This works in a finite number of steps.

\medskip
\centerline{- - - - - - - - - - - - - - - - - - - -}
\medskip

The last factor in (71) now easily satisfies
\be     \Pi_k \left( \frac{a_k}{\al_k} \right)^{\al_k/n} \ \ge e^{-\frac{1\ve}{4n}}	\ee
(if $\ve$ is small enough).  

\medskip
\centerline{- - - - - - - - - - - - - - - - - - - -}
\medskip

We are left with the task of controlling the next to last factor in (71).  We desire that

\be   \left( \frac{(\al_k (\bar n/n))!} {(\al_k\bar n/{ne})^{\al_k (\bar n/n)}} \right)^{1/\bar n} \le e^{\frac 1 M\; \frac\ve {10}} \ee
which using the estimates of (72) would follow from
\be	\frac 1 2 \; ln(2\pi) + \frac 1 2 \; ln \; \bar n + \frac 1{12} \le \frac{\bar n \ve}{10M}	\ee
where some simple steps have been omitted.

We now state our Theorem C, which will contain the conditions on parameters we've needed.

\medskip
\centerline{- - - - - - - - - - - - - - - - - - - -}
\medskip

\noindent
\underline{Theorem C}.

Given $\ve > 0$, then there are $n_1$ and $N_1$ such that if
\[	\bar n > n_1   \]
\[	N > N_1  \]
and if equations (74), (81) and (86) hold one has
\be \left| \left( Z^+(\bar f)\right)^{1/N} - \left(Z^-(\bar f)\right)^{1/N} \right| < \ve    \ee

\medskip
\centerline{- - - - - - - - - - - - - - - - - - - -}
\medskip

We need finally specify parameters that satisfy the conditions of all the theorems and lemmas.  The choices we will make are undoubtedly terrible, there must be a much better theorem with much weaker conditions.  We state the conditions in the form of a theorem

\medskip
\centerline{- - - - - - - - - - - - - - - - - - - -}
\medskip
\noindent
\underline{Conditions Theorem}.  Given $\ve > 0$, small enough, and $s > 0$, then if $\bar \ve \le \ve$ with
\begin{eqnarray}
sm(f) &=& \bar\ve^{n+1 + 2/d + (2n+3)s} \\
N \ge \bar n &\cong& \bar \ve^{\; -nd-2 - 2(n+1)sd} \\
\bar M &\cong& \bar \ve^{-nd - 2nds}
\end{eqnarray}
then
\begin{itemize}
\item[(1)] In Lemma 3, equation (31) holds with the right side of the inequality equal $\ve f$.  (We have made $\bar\ell \; sm(f)$ small in (30).)
\item[(2)] In Theorem A, equation (43) holds with the right side of the inequality equal to $\ve$.  ($\bar n$ goes to infinity by (89) easily faster than Theorem A requires.)
\item[(3)]  In Theorem B, equation (44) holds with the right side of the inequality equal to $\ve$.  (Again (89) ensures that $\bar n$ and $N$ both go to infinity faster than Theorem B requires.)
\item[(4)] In Theorem C, equation (87) holds with the right side of the inequality equal $\ve$.  (We will have to study (74), (81), and (86).)
\item[(5)]  The Main Theorem holds!
\end{itemize}
\medskip
\centerline{- - - - - - - - - - - - - - - - - - - -}
\medskip
\noindent
\underline{Note 1}.  $\bar n$ must be an integer to a dth power, and $\bar M$ is an integer, so the $\cong$ in equations (89) and (90) should be interpreted as say the closest such integers.

\medskip
\centerline{- - - - - - - - - - - - - - - - - - - -}
\medskip
\noindent
\underline{Note 2}.  The $\bar\ve^s$ factors in (88)-(90) enable us to smother some numerical factors, such as $dn$ in (30), and yield a factor $\ve$ in (11) of the Main Theorem, instead of $4\ve$ (from Lemma 3, Theorem A, Theorem B, and Theorem C errors added).  Likewise the dependence on $c_1$ of the Main Theorem is hidden in how small $\ve$ must be.

 \medskip
\centerline{- - - - - - - - - - - - - - - - - - - -}
\medskip
\noindent
\underline{Note 3}.  For any fixed $\La$ the limit of the Main Theorem requires only
\be	sm(f) \longrightarrow 0 .	\ee
Therefore restrictions that $N$ be large enough, such as in Theorem C, do not get reflected in the conditions of the Main Theorem.

\medskip
\centerline{- - - - - - - - - - - - - - - - - - - -}
\medskip
\noindent
\underline{Note 4}.   The fall-off of $f$, indeed the exponential nature of the fall-off, are used in an essential way in the proof of Theorem C (in the computation in the next section).  As we've said before we think there is a better theorem without such a requirement.

The computations needed to verify the Conditions Theorem are in the next section. 

\noindent
\underline{Section VI, Calculations}.

We first address ourselves to computing equation (8).  From property 3), equation (3), since $f$ is constant when $sm(f) = 0$, we get
\be	\frac{(N-1)!}{(n-1)!(N-n)!} \ f \ = \ 1   \ee
explaining the first factor of (8).  The second factor counts the ways of dividing a set of $N$ elements into $(N/n)$ subsets, each with $n$ elements.

We now turn to considering the parameter choices given in (88)-(90) of the Conditions Theorem.  The choices were derived by considering equations (31), (74), (81), and (86), though using only the first three of these really sufficed.

We first take a look at (30), that feeds into (31), and find here upon substituting (88) - (90) that
\be	\al = dn\; \bar\ve^{1+s}.	\ee
The inequality (81) becomes upon these substitutions
\be	\bar\ve^{-nd-2-2(n+1)sd} \ \ > \ \ \frac{200 n}{\ve^2} \ \bar\ve^{-nd-2nds} \ .	\ee
Using $\ve \ge \bar\ve$ (so that (95) implies (94)) and dividing some factors this inequality becomes
\be		\bar\ve^{-2sd} \ \ \ > \ \ \ 200\; n \ .		\ee
The inequality (86) becomes under substitutions (using $\ve \ge \bar\ve$, and $M \le \bar M$ so the following implies (86)
\be	\frac 1 2 \; ln(2\pi) + \frac 1 2 \; ln \left(\bar\ve^{-nd-2-2(n+1)sd}\right) + \frac 1{12} \le \frac 1{10} \bar\ve\bar\ve^{-nd-2-2(n+1)sd} \; \bar\ve^{nd \; + \; 2nds} \ .   \ee
 Clearly (95) and (96) hold for small $\bar\ve$.  And $\al$ from (93) ensures $\frac{\al}{1-\al} \le \ve$ for small $\bar \ve$.  The proofs are complete if we can show (88) - (90) ensure (74), to which we turn.

We can identify supertypes with supersets pointed at the cube containing the origin.  Each such superset is a cube in an nd-dimensional lattice. The identification of supersets described after (49) does not change the succeeding arguments.  We consider a large sphere of radius $R$ centered at the origin (of this nd-dimensional lattice).  Let the number of the above described supersets not entirely outside the sphere be $\bar M$.  We have
\be		\bar M \ \le \ \bar c_1 \left(\frac R \ell \right)^{nd}   \ee
or
\be		R \ \ge \ \left( \frac{\bar M}{\bar c_1} \right)^{1/nd} \ \ell \ .	\ee
Remember the edge size of the cubes representing the supertypes is $\ell$.  Referring to equation (74) and recalling the definition of the $a_k$ one has
\be	\sum^{\bar M}_{k=1} a_k \ge 1 - \bar c_2 \int^\infty_R e^{-\bar c_3r\de} \ r^{nd-1} \ dr   \ee
where we have taken the $a_k$ in the sum on the left to be those of supertypes not outside the sphere, and we have clearly used (5) and (13), with $\de \equiv sm(f)$.
\be	\bar c_2 \int^\infty_R e^{-\bar c_3r\de} \ r^{nd-1} \ dr   \le e^{-\hat c\; \bar M^{1/nd}\ell\de}    \ee
when $\bar M^{1/nd}\ell\de$ is large.   $\bar M^{1/nd}\ell\de$ from the values in (88)-(90) is
\be \bar\ve^{-1-2s} \; \bar \ve^{-n-2/d \; - \; 2(n+1)s} \ \bar\ve^{n+1 \; + \; 2/d \; + \; + (2n+3)s} = \bar \ve^{-s} \ . \ee
Then turning to (100) and (99) we get
\be		\sum^{\bar M}_{k=1} a_k \ge 1 - e^{-\hat c \bar\ve^{-s}}  \ .   \ee

La commedia $\grave{e}$ finita !

\bigskip

 \medskip
\centerline{- - - - - - - - - - - - - - - - - - - -}
\medskip

\centerline{\underline{References}}

\noindent
William Feller, An Introduction to Probability Theory and Its Applications, Vol 1, John Wiley and Sons, 1950.

\end{document}